\documentclass[aps,pre,reprint,superscriptaddress]{revtex4-1}

\usepackage{graphicx}
\usepackage{times}
\usepackage{epstopdf}
\usepackage{amssymb}   % for math
\usepackage{amsfonts}
\usepackage{amsmath}
\usepackage{mathrsfs}
\usepackage{mathtools}
\usepackage{times}
\usepackage{xcolor}
\usepackage{adjustbox}

\DeclareMathOperator*{\argmin}{arg\,min}
\DeclareMathOperator*{\argmax}{arg\,max}

\newcommand{\thetas}{\mathbf{\theta}}
\newcommand{\etas}{\mathbf{\eta}}
\newcommand{\ps}{\mathbf{p}}
\newcommand{\hatthetas}{\mathbf{\hat{\theta}}}
\newcommand{\hatetas}{\mathbf{\hat{\eta}}}
\newcommand{\hatps}{\mathbf{\hat{p}}}
\newcommand{\R}{\mathbb{R}}

\begin{document}

%\preprint{}

\title{Accurate and scalable social recommendation using mixed-membership stochastic block models}


%\author{}
%\thanks{}
%\affiliation{}

\author{Antonia Godoy-Lorite}
\email{antonia.godoy@urv.cat}
%\thanks{Corresponding author}
\affiliation{Departament d'Enginyeria Qu\'{\i}mica, Universitat Rovira i Virgili, 43006 Tarragona, Catalonia}
\author{Roger Guimer\`a}
\email{roger.guimera@urv.cat}
%\thanks{Corresponding author}
\affiliation{Departament d'Enginyeria Qu\'{\i}mica, Universitat Rovira i Virgili, 43006 Tarragona, Catalonia}
\affiliation{Instituci\'o Catalana de Recerca i Estudis Avan\c{c}ats (ICREA), 08010 Barcelona, Catalonia}
\author{Cristopher Moore}
\email{moore@santafe.edu}
\affiliation{Santa Fe Institute, Santa Fe, NM 87501, USA}
\author{Marta Sales-Pardo}
\email{marta.sales@urv.cat}
\affiliation{Departament d'Enginyeria Qu\'{\i}mica, Universitat Rovira i Virgili, 43006 Tarragona, Catalonia}


\date{\today}

\begin{abstract}
With ever-increasing amounts of online information available, modeling and predicting individual preferences---for books or articles, for example---is becoming more and more important. Good predictions enable us to improve advice to users, and obtain a better understanding of the socio-psychological processes that determine those preferences. We have developed a collaborative filtering model, with an associated scalable algorithm, that makes accurate predictions of individuals' preferences. Our approach is based on the explicit assumption that there are groups of individuals and of items, and that the preferences of an individual for an item are determined only by their group memberships. Importantly, we allow each individual and each item to belong simultaneously to mixtures of different groups and, unlike many popular approaches, 
such as matrix factorization, 
we do not assume implicitly or explicitly that individuals in each group prefer items in a single group of items. The resulting overlapping groups and the predicted preferences can be inferred with a expectation-maximization algorithm whose running time scales linearly (per iteration) 
% CM: added this
with the number of observed ratings. Our approach enables us to predict individual preferences in large datasets, and is considerably more accurate than the current algorithms for such large datasets.
\end{abstract}

%\keywords{}

\maketitle

The goal of recommender systems is to predict what movies we are going to like, what books we are going to purchase, or even who we might be interested in dating. The rapidly growing amount of data on item reviews, ratings, and purchases from a growing number of online platforms holds the promise to facilitate the development of finer and more informed models for recommendation. At the same time, however, it poses the challenge of developing algorithms that can handle such large amounts of data both accurately and efficiently.

%Importantly, besides the obvious economic value associated to good recommendation
%engines, these also have a value from a fundamental point of view; recommender systems
%based on sound assumptions can also provide good models to reproduce and understand the
%the socio-psychological processes that determine user preferences. 

A plausible expectation when developing recommendation 
%models and 
algorithms is that similar users relate to similar objects in a similar manner, i.e., they purchase similar items and give the same item similar ratings. This means that we can use the rating history of a set of users to make recommendations, even without knowing anything about the characteristics of users or items; this is 
%precisely 
the basic underlying assumption of collaborative filtering, one of the simplest and most common approaches in recommender systems~\cite{su09}.  However, most research in recommender systems has not focused on precisely formalizing these general assumptions into plausible and rigorous models, but rather on the development of scalable algorithms, often at the price of implicitly using models that are overly simplistic or unrealistic. For example, matrix factorization and latent feature approaches assume that users and items live in some abstract low-dimensional space, but whether such a space is expressive enough to accommodate for the rich variety of user behaviors is rarely discussed. As a result, such state-of-the-art scalable approaches have significantly lower accuracies than inference approaches based on models of user preferences that are socially more realistic~\cite{guimera12}. On the other hand, these more realistic approaches do not scale well with dataset size, which makes them unpractical for large datasets. 

Here, we develop an approach to predict user ratings that makes explicit hypotheses about rating behavior. In particular, our approach is based on the assumption that there are groups of users and of items, and that the rating a given user assigns to a given item is determined probabilistically by their group memberships. Importantly, we do not assign users and items to a specific group; rather, we allow each user and each item to belong simultaneously to mixtures of different groups~\cite{airoldi08,peixoto15}. All of these elements are combined in a model with a precise probabilistic interpretation, which allows for rigorous inference algorithms.  Happily, the inference problem for our model can be solved very efficiently: specifically, we propose an expectation-maximization algorithm whose running time, per iteration, scales linearly with the number of observed ratings, and which appears to converge rapidly in practice.

We demonstrate that our model is more realistic than those implicit in other approaches (particularly matrix factorization) and that, as a consequence, our approach consistently outperforms state-of-the-art collaborative filtering approaches, often by a large margin. Moreover, because our model has a clear interpretation, it can deal naturally with some situations that are challenging for other approaches (for example, the cold start problem) and can help to build theories about user behavior. We argue that our approach may also be suitable for other areas where matrix factorization is increasingly used such as image reconstruction, textual data mining, cluster analysis or pattern discovery \cite{cemgil09,berry07,ding05,kim08,brunet04}.

\section{A mixed-membership block model with metadata}

Our approach begins with the mixed-membership stochastic block model (MMSBM), which has been used to model networks with overlapping communities or groups.  As in the original MMSBM~\cite{airoldi08} and in related models~\cite{ball-karrer-newman}, we assume that each node in the bipartite graph of users and items belongs to a mixture of groups.  However, unlike in~\cite{airoldi08,ball-karrer-newman}, we do not assume that these group memberships affect the presence or absence of an link, i.e., the event that a given user rates a given item.  Instead, we take the set of links as given, and attempt to predict the ratings.  We do this with an MMSBM-like model where the rating a user gives an item is drawn from a probability distribution that depends on their group memberships.  

Let us set down some notation.  We have $N$ users and $M$ items, and a bipartite graph $R = \{ (u,i) \}$ of links, where the link $(u,i)$ indicates that item $i$ was given a rating (observed or unobserved)  by user $u$.  For each $(u,i) \in R$, the rating $r_{ui}$ belongs to some finite set $S$ such as $\{1,2,3,4,5\}$.  Given a set $R^O$ of observed ratings, our goal is to classify the users and the items, and to predict the rating $r_{ui}$ of a link $(u,i) \in R$ for which the rating is not yet known.

Our generative model for the ratings is as follows.  There are $K$ groups of users and $L$ groups of items.  For each pair of groups $k,\ell$, there is a probability distribution $p_{k\ell}(r)$ over $S$ of the rating $r$ that $u$ gives $i$, assuming that $u$ belongs entirely to group $k$ and $i$ belongs entirely to group $\ell$.  

To model mixed group memberships, each user $u$ has a vector $\theta_u \in \R^K$, where $\theta_{uk}$ denotes the extent to which user $u$ belongs to group $k$.  Similarly, each item $i$ has a vector $\eta_i \in \R^L$.  These vectors are normalized, i.e., $\sum_k \theta_{uk} = \sum_\ell \eta_{i\ell} = 1$.  Given $\theta_u$ and $\eta_i$, the probability distribution of the rating $r_{ui}$ is then a convex combination,
\begin{equation}
 \Pr[r_{ui} = r] = \sum_{k,\ell} \theta_{uk} \eta_{i\ell} p_{k\ell}(r) \, .
 \label{eq:model}
\end{equation}
%
%This model is the bipartite version of the mixed-membership stochastic block model (MMSBM) for networks in Ref.~\cite{airoldi08}, or the mixed-membership version of the bipartite stochastic block model (SBM) that has been used before as a recommender system \cite{guimera12}.
Abbreviating all these parameters as $\thetas, \etas, \ps$, the likelihood of the observed ratings is thus 
\begin{equation}
P(R^O | \thetas, \etas, \ps) = \prod_{(u,i) \in R^O} \sum_{k,\ell} \theta_{uk} \eta_{i\ell} p_{k\ell}(r_{ui}) \, . 
\label{eq:likeli}
\end{equation}
As we discuss below, we infer the values of the parameters $\hatthetas, \hatetas, \hatps$ that maximize this likelihood using an efficient expectation-maximization algorithm.  We can then use the inferred model to predict unobserved ratings $r_{ui}$.

Our work is different from previous work on collaborative filtering in several ways. First, unlike matrix factorization approaches such as~\cite{koren09} or their probabilistic counterparts~\cite{meeds06,salakhutdinov08,shan10}, we do not think of the ratings $r_{ui} \in \{1,2,3,4,5\}$ as integers. As has been established in the literature, giving a movie a rating of 5 instead of 1 does not mean the user likes it five times as much~\cite{ekstrand11}.  Our results suggest that it is better to think of different ratings simply as different labels that appear on the links of the network.  
%An additional advantage of our approach is that
Moreover, our method yields a distribution over the possible ratings directly, 
%and only those, 
rather than a distribution over integers or reals that must be somehow mapped to the space of possible ratings~\cite{meeds06,salakhutdinov08,shan10}. From this point of view, our model is a bipartite MMSBM with metadata (or labels) on the edges; a similar model based on the stochastic block model (SBM), where each user and item belongs to only one group, was given in~\cite{guimera12}. An alternative approach would be to consider a multi-layer representation of the data as in~\cite{peixoto15}.
% CM: some rearrangement here.

Second, we do not assume that the matrices $\ps$ have any particular structure. In particular, we do not assume homophily, where groups of individuals correspond to groups of items, and individuals prefer items that belong to their own group: that is, we do not assume that $\ps(r)$ is larger on the diagonal for higher ratings $r$. Thus our model, and our algorithm, can learn arbitrary couplings between groups of individuals and groups of items, and do so independently for each possible rating.

Third, unlike some approaches that use inference methods similar to ours~\cite{gopalan13}, and as stated above, our goal is not to predict the \emph{existence} of links.  In particular, we do not assume that individuals only see movies (say) that they like, and we do not treat missing links as zeroes or low ratings.  To put this differently, we are not trying to complete $R$ to a full matrix of ratings, but only to predict the unobserved ratings in $R \setminus R^O$.  Thus the only terms in the likelihood of our model correspond to observed ratings.

As we describe below, our model also has the advantage of being mathematically tractable.  It yields an expectation-maximization algorithm for fitting the parameters which is highly efficient: each iteration takes linear time as a function of the number of users, items, and observed links.  As a result, we are able to handle quite large datasets, and achieve a higher accuracy than standard methods.

%--------------------------------------------------------------------------THE MODEL
\section{Scalable inference of model parameters}

%Given the considerations above, one should expect recommender systems based on stochastic block models to be more accurate than those based on matrix factorization. Indeed, results in Ref.~\cite{guimera12} suggest that that is the case. Unfortunately, the current approach to making recommendations based on stochastic block models relies on Markov chain Monte Carlo sampling, and therefore does not scale to large datasets \cite{guimera12}; in practice, the approach is therefore of limited interest. Here, we develop an inference algorithm for mixed-membership stochastic block models that scales with the number of observed ratings and can be applied to datasets with tens of millions of ratings (and can be trivially parallelized for even larger datasets).

%Given the probability of a given rating in Eq.~\eqref{eq:model} and a set $R$ of observed ratings (the training set), the log-likelihood $\mathscr{L}$ of the model is
%
%\begin{equation}
%\mathscr{L}=\log P(R| \thetas, \etas, \ps)
%= \sum_{(u,i)\in R} \log \left( \sum_{k\ell}\theta_{uk} p_{k\ell}^{r_{ui}} \eta_{i\ell}  \right) \, 
%\end{equation}
%
%where the first summation runs over the observed ratings.

In most practical situations, marginalizing exactly over the group membership vectors $\thetas$ and $\etas$ and the probability matrices $\ps$ (similar to Ref.~\cite{guimera12}) is too computationally expensive. As an alternative we propose to obtain the model parameters that maximize the likelihood~\eqref{eq:likeli} using an expectation-maximization (EM) algorithm.
%, and then use those parameters to estimate unobserved ratings.

In particular, we use a classic variational approach (see Methods) to obtain the following equations for the model parameters that maximize the likelihood, 
\begin{eqnarray}
  \theta_{uk} = \frac{\sum_{i \in \partial u} \sum_l \omega_{ui}(k,\ell)}{d_u} \, , \label{eq:upd-theta} \\
  \eta_{i\ell} = \frac{\sum_{u \in \partial i} \sum_k \omega_{ui}(k,\ell)}{d_i} \, , \label{eq:upd-eta} \\
  p_{k\ell}(r) = 
  \frac{\sum_{(u, i) \in R^O | r_{ui}=r} \omega_{ui}(k,\ell)}
  {\sum_{(u, i) \in R^O} \omega_{ui}(k,\ell)} \, .
  \label{eq:upd-pr}
\end{eqnarray}
Here $\partial u = \{ i | (u,i) \in R^O \}$ and $\partial i = \{ u | (u,i) \in R^O \}$ denote the neighborhoods of $u$ and $i$ respectively; $d_u = |\partial u|$ and $d_i = |\partial i|$ are the node degrees, i.e., the number of observed ratings for user $u$ and item $i$ respectively; and
\begin{equation}
  \omega_{ui}(k,\ell) = \frac{\theta_{uk} \eta_{i\ell} p_{k\ell}(r_{ui})}{\sum_{k',\ell'} \theta_{uk'}  \eta_{i\ell'} p_{k'\ell'}(r_{ui}) } 
  \label{eq:upd-omega}
\end{equation}
is the variational method's estimate of the probability that the rating $r_{ui}$ is due to $u$ and $i$ belonging to groups $k$ and $\ell$ respectively.
% CM: added this

These equations can be solved iteratively with an EM algorithm.  Starting with an initial estimate of $\thetas$, $\etas$, and $\ps$, we repeat the following steps until the parameters converge:
\begin{enumerate}
\item (Expectation step) use~\eqref{eq:upd-omega} to compute $\omega_{ui}(k,\ell)$ for $(u,i) \in R^O$,
\item (Maximization step) use~\eqref{eq:upd-theta}-\eqref{eq:upd-pr} to compute $\thetas$, $\etas$, and $\ps$.
\end{enumerate}
The number of parameters and terms in the sums in Eqs.~\eqref{eq:upd-theta}-\eqref{eq:upd-omega} is $NK+ML+|R^O| KL$.  Assuming that $K$ and $L$ are constant, this is $O(N+M+|R^O|)$, and hence linear in the size of the dataset (see Fig.~S1 in Supplementary Materials (SM)).  As the set of observed ratings $R^O$ is typically very sparse because only a small fraction of all possible user-item pairs have observed ratings, our algorithm is feasible even for very large datasets.

%In summary, our MMSBM approach has a double advantage: (i) it uses a model that is more realistic and flexible than the model underlying matrix factorization approaches; (ii) the algorithm scales with the number of observed ratings, and is therefore suitable for very large datasets.
%In addition, it is consistent for sparse datasets, giving good results with few ratings per user (10 ratings per user are enought to give a good prediction).

%We develope an extension of the MMS, the Hierarchical Mixed-Membership Stochastic Blockmodel (HMMSB)(see Methods), that presents an improvement in terms of accuracy for the prediction \ref{ml5}, but more sensible for those particular ratings with less statistics. In particular, it is very useful for those datasets where the proportion of the different ratings are uneven, then those ratings with less statistics they are camouflaged in the classification (\ref{confusion}). In this algorithm there is a doble classification, first into big categories, and then a refined classification. The scaling properties are equal the previous MMSBM, being the computational time lineal with the corpus size. 

\section{Results}

\subsection{The MMSBM predicts ratings accurately} \ 
We test the performance of our algorithm by considering six datasets: the MovieLens 100K and 10M datasets with 100,000 and 10,000,000 ratings respectively, Yahoo! songs,  Amazon books~\cite{mcAuley15,mcAuley15a}, and the dataset from LibimSeTi.cz dating agency~\cite{brozovsky07}, which we split into two datasets, consisting of males rating females and vice versa. These datasets are diverse in the types of items considered, the sizes $|S|$ of the sets of possible ratings, and the density of observed ratings (see Table~\ref{t-data}).  For each dataset we perform a five-fold cross-validation, splitting it into five equal subsets, and using each one as a test set after training the model on the union of the other four.
% CM: rephrased this

We compare our algorithm to three benchmark algorithms (see Methods): a baseline naive algorithm that assigns to each test rating $r_{ui}$ the average of the observed ratings for item $i$; the item-item algorithm, which predicts $r_{ui}$ based on the observed ratings of user $u$ for items that are the most similar to $i$; and ``classical'' matrix factorization~\cite{koren09,gopalan13}. For all these benchmark algorithms we use the implementation in the LensKit package~\cite{ekstrand11}. Additionally, for the smallest datasets, we also use the (un-mixed) stochastic block model approach of Ref.~\cite{guimera12}; however, that algorithm does not scale well to larger datasets.

For our algorithm, we set $K=L=10$, i.e., we assume that there are 10 groups of users and 10 groups of items (recall that we do not assume any correspondence between these groups).  We considered some other choices of $K$ and $L$ as well (see Fig.~S2 in the SM). Since iterating the EM equation of Eqs.~\eqref{eq:upd-theta}-\eqref{eq:upd-omega} can lead to different solutions depending on the initial conditions, we perform sampling of 500 independent runs with random initial conditions. We average the predicted probabilities over the 500 runs because we typically do not observe that one solution has much higher likelihood than the others (see Fig.~S3 of the SM for results obtained using the maximum likelihood solution). As a result, for each rating a user gives an item we have a probability distribution of ratings that results from the average of the probabilities for all the sampling set. Therefore, we can choose how to make predictions from the probability distribution of ratings: the most likely rating, the mean or the median. In contrast, recommender systems like MF and item-item give only the most probable rating. We measure the performance in terms of accuracy, i.e., the fraction of ratings that are exactly predicted by each algorithm, and the mean absolute error (MAE). For our algorithm, we find that the best estimator for the accuracy is the most likely rating from the probability distribution of ratings, while for the MAE the best estimator is the median. 

\begin{figure}
  \includegraphics*[width=\columnwidth]{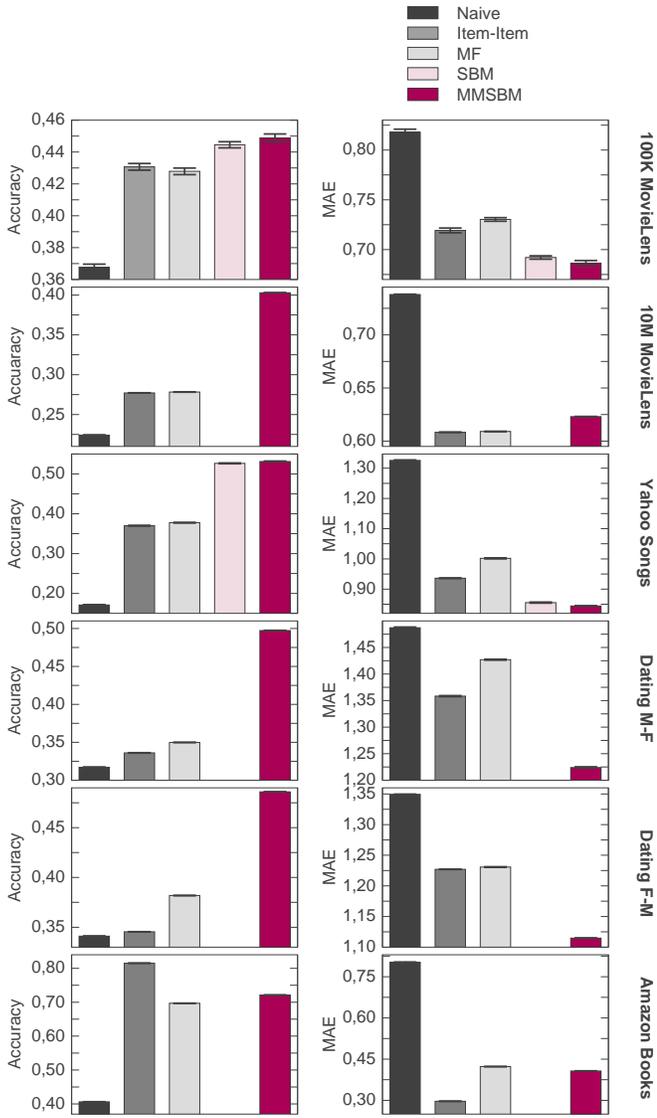}
  \vspace{4mm}
\caption{Algorithm comparison. From top to bottom, the datasets are
MovieLens 100K,  % (with ratings $S=\{1,\ldots,5\}$), 
Movielens 10M, %(where $S=\{0.5,1,1.5,\ldots,5\}$), 
Yahoo Songs,  %(311,700 ratings, $S=\{1,\ldots,5\}$), 
men rating women (M-W) in the LibimSeTi dataset, % (4,852,455 ratings, $S=\{1,\ldots,10\}$), 
women rating men (W-M)  in the LibimSeTi dataset %(10,804,040 ratings, $S=\{1,\ldots,10\}$) 
and Amazon books. %(4,505,893 ratings, $S=\{1,\ldots,5\}$). 
The left column displays the accuracy of the algorithms in each dataset, i.e., the fraction of ratings that are exactly predicted by each algorithm.  The right column displays the mean absolute error (MAE) in the predicted vs.\ actual rating, treated as an integer or half-integer.  In all cases, the bars are the average of a five-fold cross-validation and the error bars correspond to the standard error of the mean.  The SBM algorithm does not scale to the larger datasets, but achieves similar accuracy to the MMSBM on the datasets it can handle.  The MMSBM model and algorithm of this paper achieves the best (highest) accuracy in five out of size datasets, and the best (lowest) MAE in four out of six datasets.
\label{f.compare}}
\end{figure}

We find that in most cases our approach outperforms the item-item algorithm and matrix factorization (Fig.~\ref{f.compare}). Indeed, when considering the accuracy, i.e., the fraction of times an algorithm exactly predicts the correct rating, the MMSBM is significantly better than matrix factorization for all the datasets we tested, and better than the item-item algorithm in five out of  six datasets, the only exception being the Amazon Books dataset. In terms of the mean absolute error (MAE), the MMSBM is the most accurate in four out of the six datasets (item-item and matrix factorization produce smaller MAE in the Amazon Books and MovieLens 10M datasets). \footnote{Note that the Amazon dataset is different from the others in that users only rate items after buying them, and know a priori the average rating of the item given by previous buyers, which might bias their choices.}

Interestingly, our approach produces results that are almost identical to those of the un-mixed SBM~\cite{guimera12} for the two examples for which inference with the SBM is feasible. 
In particular, we achieve the same accuracy with $K=L=10$ in the mixed-membership model as with around $50$ groups in the un-mixed SBM.  This suggests that many of the groups observed in~\cite{guimera12} are in fact mixtures of a smaller number of groups, and that the additional expressiveness of the MMSBM allows us to succeed with a lower-dimensional model.
%of SBMs and MMSBMs is responsible for the high accuracy of these approaches, and that the effect of averaging over user and item partitions as in Ref.~\cite{guimera12} is somewhat equivalent to assuming mixed membership.  
% CM: changed this

\subsection{MMSBMs generalize matrix factorization and provide more expressive models} 
\begin{figure*}
\centerline{\includegraphics*[width=2\columnwidth]{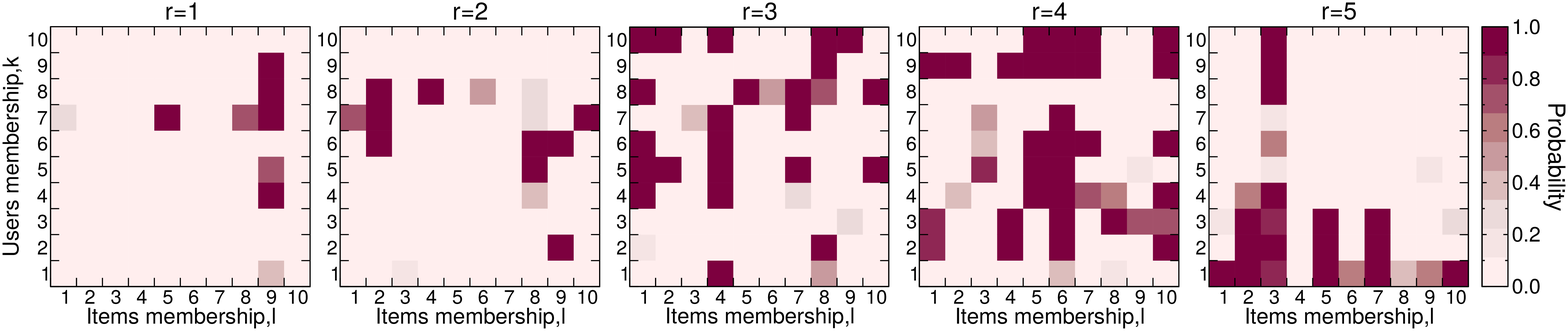}}
\vspace{-3.2cm}
\caption{The inferred values for the probability matrices $\ps$ from the MovieLens 100K dataset.  Left to right, the five matrices correspond to the ratings $r=1,2,3,4,5$.  For each one them, the rows and columns correspond to the user's and item's groups; here $K=L=10$.  Each element, shown as a heat map, gives the probability $p_{k\ell}(r)$ that a user in group $k$ gives a rating $r$ to an item in group $\ell$. The matrices are normalized as shown in~\eqref{eq:norm-pr}.  Notice that there is no ordering of the probability matrices that would make them diagonal.
\label{f.pmatrices}}
\end{figure*}

Matrix factorization (MF) is one of the most successful and popular approaches to collaborative filtering, both in its ``classical''~\cite{koren09} and its probabilistic form~\cite{meeds06,salakhutdinov08,shan10,gopalan13}. However, as we have just discussed, our 
%scalable 
MMSBM gives consistently more accurate results for the ratings, often by a large margin. Here, we analyze the origin of this improvement in performance.

We start by giving an interpretation of matrix factorization in terms of our MMSBM. A matrix is of rank $K$ if and only if its entries can be written as inner products of $K$-dimensional vectors associated with its rows as columns. Based on this idea, matrix factorization assumes that the expected rating that user $u$ gives item $i$ is $\bar{r}_{ui} = \tilde{\theta}_u \cdot \tilde{\eta}_i$, where $\tilde{\theta}_u$ and $\tilde{\eta}_i$ are $K$-dimensional vectors 
%(with $K$ much smaller than the number of users $N$ and the number of items $M$) 
representing the user and the item respectively. One can apply a variety of noise models or loss functions, as well as regularization terms for the model parameters~\cite{koren09}, but this does not alter significantly the considerations that we present next.
%.  (Most often, the model also includes user and item biases as well as a regularization term 

%Since users and items are represented as vectors  in a $K$-dimensional space, matrix factorization can be interpreted geometrically---in general, the model predicts that users will like items that are well aligned with them, and only those; and that if two users are not aligned with each other there will be no items that they will both like. These features appear as somewhat unrealistic, and we argue that they limit the performance of matrix factorization.

The limitations in expressiveness of matrix factorization become apparent when we interpret matrix factorization as a mixture model.  Assume that there are $K$ groups of users and that $\theta_{uk}$ is the probability that user $u$ belongs to group $k$. Similarly, assume that there are $K$ groups of items and that $\eta_{ik}$ is the probability that item $i$ belongs to group $k$. Finally, assume that users in group $k$ \emph{only} like items in group $k$; in particular, users in $k$ assign a baseline rating of $1$ to items in group $k$ and a rating of $0$ to items in all other groups.  Finally, let $s_u \ge 0$ and $s_i \ge 0$ be user and item ``intensities'' that correct for the fact that some users rate on average higher than others, and that some items are generally more popular than others. Then the expected ratings are given by %of user $u$ for item $i$ is
\begin{equation}
\label{eq:mf-limited}
 \bar{r}_{ui} = \sum_k s_u \theta_{uk} s_i \eta_{ik} \, .
\end{equation}
Identifying $\tilde{\theta}_{uk} = s_u \theta_{uk}$ and $\tilde{\eta}_{ik} = s_i \theta_{ik}$, this becomes the matrix factorization model $\bar{r}_{ui} = \tilde{\theta}_u \cdot \tilde{\eta}_i$.  %Conversely, any rank $K$ matrix which can be written as a sum of $K$ rank-one matrices with nonnegative entries can be written in the form~\eqref{eq:mf-limited}. 
Thus (nonnegative) matrix factorization corresponds to a model where each group of users corresponds to a group of items, and users in a given group only like items in the corresponding group. We argue that these assumptions are too limiting to model user recommendations realistically. (Note that our interpretation of matrix factorization as a mixture model is independent of attempts in the literature to combine matrix factorization with other mixture models \cite{mackey10}.)

Our MMSBM relaxes these implausible assumptions by allowing the distribution of ratings to be given by arbitrary matrices $\ps$, where the entry $p_{k\ell}(r)$ is the probability that a user in group $k$ gives an item in group $\ell$ the rating $r$.  Matrix factorization is roughly equivalent to assuming that $p_{k\ell}$ is diagonal, at least for high ratings.  
We believe that the improved performance of the MMSBM over matrix factorization is due to this greater expressive power. Indeed, 
%we can check if the introduction of the rating probability matrices is responsible for the improved performance. To this end, we analyze the $\ps$ matrices that maximize the likelihood of the MMSBM (Fig.~\ref{f.pmatrices}). We observe that, indeed, 
Fig.~\ref{f.pmatrices} shows that the matrices $\ps$ inferred by our model are far from the purely diagonal structure implicitly underlying matrix factorization.

Moreover, the generality of the MMSBM allows it to account for many of the features of real ratings.  For instance, the distribution of ratings is highly nonuniform: as shown in Fig.~\ref{f.pmatrices}, $r=1$ is quite rare whereas $r=4$ is quite common.  Different groups of users have very different distributions of ratings: users in group $k=1$ rate most movies with $r=5$, while those in group $k=7$ often give ratings $r=1$.   Similarly, movies in group $\ell=3$ are consistently rated $r=5$ by most users, while movies in group $\ell=9$ are rated $r=1$ quite often.  It is also interesting that some groups of users agree on some movies but disagree on others: for example, users in groups $k=9, 10$ agree that most movies in group 
%$\ell=8$ should be rated $r=3$, 
% CM: changed example here
$\ell=3$ should be rated $r=5$, but they disagree on movies in group $\ell=9$, rating them $r=1$ and $r=3$ respectively. 
%) and some groups of movies are consistently given the same rating by all users (for example, movies in $l=3$ are consistently given $r=5$ by most users); some users agree on rating a group of movies, while disagreeing on others (for example, users in $k=9$ and $k=10$ agree at rating with $r=3$ movies in $l=8$, but users in $k=9$ consistently rate movies in $l=9$ with $r=1$, whereas users in $k=10$ consistently rate the same movies with $r=3$). 
These observations highlight the limitation in expressiveness of matrix factorization, and explain why our approach based on MMSBM yields better predictions of the ratings.

\subsection{The MMSBM provides a principled method to deal with the cold start problem}
Because in the MMSBM all terms have a clear and precise (probabilistic) interpretation, our approach can naturally deal with situations that are challenging for other algorithms. An example of this is the cold start problem, that is, a situation in which we want to predict ratings for users or items (or both) for which we do not have training data \cite{schein02,salakhutdinov08,park09}.

In the MMSBM, the $\ps$ matrices are the same for all users and items; in this sense, new users or items pose no particular difficulty. However, for a new user $n$ we need to calculate their group membership vector $\theta_n$ (and analogously $\eta_i$ for a new item).  
%; here we focus on users but the discussion for items is analogous). 
Since on average users tend to have a higher probability of belonging to some groups than to others, lacking all information about a user we can assume that they are proportionally more likely to belong to the same groups. In practice, this means that to any new user $n$ we can assign a group membership vector that is the average of the vectors of the observed users,
% CM: changed "must" to "can" here
%
\begin{equation}
 \theta_{nk} = \frac{1}{N} \sum_u \theta_{uk} \;.
\end{equation}
This provides a principle method to deal with the cold start problem, without the need to add additional elements to the model~\cite{salakhutdinov08}.

\begin{figure}
\centerline{\includegraphics*[width=\columnwidth]{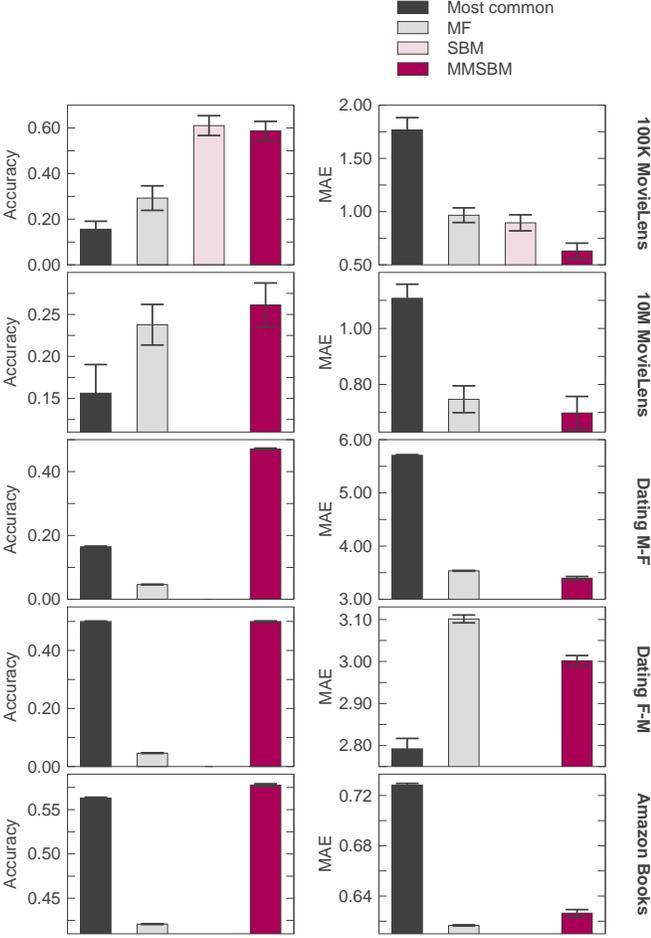}}
\caption{Algorithm performance for the cold start problem. From top to bottom: the MovieLens 100K dataset 
%(ratings scale from 1 to 5, precision 1) 
with $0.17\%$ of cold start cases on average; the Movielens 10M dataset 
%(ratings scale from 1 to 5, precision 0.5) 
($0.0015\%$); men rating women (M-W) in the LibimSeTi dataset 
%(ratings scale from 1 to 10, precision 1) 
($0.625\%$); women rating men (W-M) in the LibimSeTi dataset 
%(ratings scale from 1 to 10, precision 1) 
($0.31\%$); and Amazon books 
%(ratings scale from 1 to 5, precision 1) 
($6.7\%$).  We did not encounter any cold start cases in the cross-validation experiments with Yahoo! Songs; this is to be expected since Yahoo! Songs requires that users and songs have at least 20 ratings.
%, it is statistically unlikely to find cold start cases in a cross-validation experiment by hiding 20\% of the data. 
The left column displays the accuracy for each dataset, and the right column the mean absolute error. The bars show the average of a five-fold cross-validation and the error bars show the standard error of the mean.
\label{f.coldstart}}
\end{figure}

In Fig.~\ref{f.coldstart} we show that, also in cold start situations, our MMSBM outperforms the alternatives in most cases. In terms of accuracy, MMSBM is always more accurate than MF (although in one case the difference is not significant), and more accurate than just assigning the most common rating to an item in all cases but one. In terms of mean absolute error, our approach is more accurate than MF in four out of five cases (in one, not significantly), and more accurate than using the most common rating in four out of five cases. 

\subsection{Groups inferred with the MMSBM reflect features of users} 
Finally, the expressiveness of the MMSBM enables us to investigate the social and psychological processes that determine user behaviors. To illustrate this idea, we analyze the user profiles in the MovieLens 100K dataset, which lists the age and gender of each user. 

Specifically, we compare the user profiles of pairs of users $(u,v)$ by computing the cosine similarity $\sum_k \theta_{uk}\theta_{vk}/(|\theta_u|_2 |\theta_v|_2)$.
%
%A plausible expectation is that groups of users with similar attributes have larger similarity.
%
\begin{figure}
\centerline{\includegraphics*[width=\columnwidth]{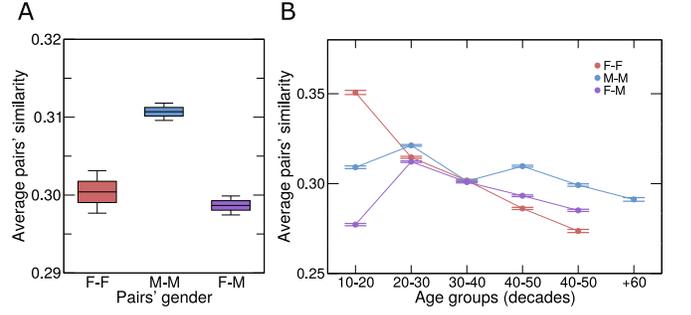}}
\caption{User profile similarities in the MovieLens 100K dataset by gender and age.  For each pair of users $(u,v)$, we compute the cosine similarity of their user profiles $\sum_k \theta_{uk}\theta_{vk}/(|\theta_u|_2 |\theta_v|_2)$. Panel A shows the average similarity for pairs of females (F-F), pairs of males (M-M) and mixed gender pairs (F-M).  The boxes show the mean (black line) and one standard error of the mean; the bars show two standard errors of the mean. Panel B shows average user similarities among users in the same age group, as a function of age.  Note that there are no female users of age greater than $60$. 
%We show the average similarity of age pairs of users in the same age group (bars show one standard deviation of the mean) for each category. 
The data suggests that male users are slightly more similar to each other than female users are, and that for all gender pairs similarity decreases with age (F-F: Spearman's $\rho=-0.078$; p-value$=2.34\cdot 10^{-24}$; M-M: 
Spearman's $\rho=-0.020$, p-value$=1.24\cdot 10^{-10}$; Spearman's $\rho=-0.016$, p-value$=4.58\cdot 10^{-6}$).}
\label{f.agesex}
\end{figure}

Figure~\ref{f.agesex} shows that when when we divide users according to gender, pairs of male users have more similar profiles than pairs of female users or male-female pairs (see Fig.~\ref{f.agesex}A). Interestingly, when we combine gender and age to define user groups, we find that gender profile similarities are not independent of the age groups (see Fig.~\ref{f.agesex}B). In fact, we observe the general tendency that young users within a gender group seem to have larger profile similarities than older users. Interestingly, this tendency is more apparent for female users who are the group with larger similarity for ages 10-20 and the one with lower similarity for ages 40-50.

\section{Discussion}
Our results show that the MMSBM we propose, and its associated expectation-maximization algorithm, is a robust, well-performing, and scalable solution to predict user-item ratings in different contexts. Additionally, the interpretability of its parameters enables the analysis of the underlying social behavior of users.  For example, we found that the similarity of users' behavior is correlated with their gender and their age.  
%we have shown that 
%younger users display rating patterns that are more similar than those of older users, and that this feature is further exacerbated in female users. 
% CM: I'm unwilling to make this one of the main messages of the paper without a lot more analysis.
% CM: Also, note that matrix factorization would also allow one to compute cosine similarity. 
These findings could conceivably lead to extensions of the model that take such behavioral considerations into account, for example by adding metadata to users (e.g. age and gender) and items (e.g. genre). In fact, stochastic block models with node metadata have recently been proposed~\cite{newman15} and may be a promising way to extend our approach.

%, suggesting that the disparity in rating patterns is more accentuated in middle aged female users than in male users in the same age bracket.
Another advantage of the interpretability of our model and its parameters is that it can be readily applied to (and performs well in) situations that are challenging to other approaches, such as a cold start where no prior information is available about a new user or item.  
%For the same reason, it may feasible to extend our approach to other situations where traditional approaches do not perform well or simply cannot be applied.
% CM: this sentence is too vague.

%, one of the hard issues in collaborative system approaches. While it is not our goal to compare our algorithm to content based approaches that have better performances in cold start cases, our approach is a promising alternative to build a hybrid recommendation system that incorporates contextual information.

Finally, 
%it is remarkable that 
the MMSBM outperforms matrix factorization in all the cases we consider, often by a large amount. As we have discussed, this is due to the fact that MMSBM is a more expressive generalization of the model underlying matrix factorization; matrix factorization corresponds roughly to the special case of MMSBM where the matrices $p_{k\ell}$ are diagonal, and where we assume the rating probabilities $p_{k\ell}(r)$ for different $r$ are strongly correlated (corresponding to treating $r$ as a number rather than a symbol).  
% CM: rephrased here
Matrix factorization is a widely used tool with many applications beyond recommender systems; 
%its many formats (non-negative factorization, probabilistic factorization, singular value decomposition) in a number of different applications.
%Since the MMSBM is mathematically very similar to the matrix factorization formulation,
given our findings and the scalable expectation-maximization algorithm, it may make sense to use MMSBMs in those other applications as well.

% -----------------------------------------------------------------------------------
% METHODS
% -----------------------------------------------------------------------------------
\appendix
\section{Update equations}
In the MMSBM, each user $u$ has a vector $\theta_{uk}$ describing how much she belongs to group $k$, and each item $i$ has a vector $\eta_{i\ell}$ describing how much it belongs to group $\ell$.  We treat these as probabilities, and normalize them as
\begin{equation}
\label{eq:norm-pm}
\forall u: \sum_{k=1}^K \theta_{uk} = 1 \, ,
\quad \forall i: \sum_{\ell=1}^L \eta_{i\ell} = 1 \, .
\end{equation}
Similarly, the matrices $p_{k\ell}(r)$ are normalized to give probability distributions of ratings over $S=\{1,2,3,4,5\}$, 
\begin{equation}
\label{eq:norm-pr}
\forall k,\ell: \sum_{r \in S} p_{k\ell}(r) = 1 \, .
\end{equation}
We maximize the likelihood~\eqref{eq:likeli} as a function of $\thetas, \etas, \ps$ using an expectation maximization (EM) algorithm.  We start with a standard variational trick that changes the log of a sum into a sum of logs, writing
\begin{align}
\log P(R^O | \thetas, \etas, \ps ) 
&= \sum_{(u,i) \in R^O} \log \sum_{k\ell} \theta_{uk} \eta_{i\ell} p_{k\ell}(r_{ui}) \nonumber \\
&= \sum_{(u,i) \in R^O} \log \sum_{k\ell} \omega_{ui}(k,\ell) \,\frac{\theta_{uk} \eta_{i\ell} p_{k\ell}(r_{ui})}{\omega_{ui}(k,\ell)} 
\nonumber \\
&\ge \sum_{(u,i) \in R^O} \sum_{k\ell} \omega_{ui}(k,\ell) \log \frac{\theta_{uk} \eta_{i\ell} p_{k\ell}(r_{ui})}{\omega_{ui}(k,\ell)} \, . 
\label{eq:finalL}
\end{align}
Here $\omega_{ui}(k,\ell)$ is the estimated probability that a given ranking $r_{ui}$ is due to $u$ and $i$ belonging to groups $k$ and $\ell$ respectively, and the lower bound in the third line is Jensen's inequality $\log \bar{x} \ge \overline{\log x}$. This lower bound holds with equality when 
\begin{equation}
\label{eq:update-omega}
\omega_{ui}(k,\ell) 
= \frac{\theta_{uk} \eta_{i\ell} p_{k\ell}(r_{ui})}{\sum_{k'\ell'} \theta_{uk'} \eta_{\ell 'i} p_{k'\ell'}(r_{ui})} \, , 
\end{equation}
giving us the update equation~\eqref{eq:upd-omega} for the expectation step.

For the maximization step, we derive update equations for the paremeters $\thetas, \etas, \ps$ by taken derivatives of the log-likelihood~\eqref{eq:finalL}.  Including Lagrange multipliers for the normalization constraints~\eqref{eq:norm-pm}, we obtain 
\begin{equation}
\label{eq:update-theta}
\theta_{uk} = \frac{\sum_{i \in \partial u} \sum_l \omega_{ui}(k,\ell)}
{\sum_{i \in \partial u} \sum_{k\ell} \omega_{ui}(k,\ell)} 
= \frac{\sum_{i \in \partial u} \sum_l \omega_{ui}(k,\ell)}{d_u} \, ,
%\mid (u,i) \in R 
\end{equation}
where $d_u$ is the degree of the user $u$.  Similarly, 
\begin{equation*}
%\label{eq:update-eta}
\eta_{i\ell} = \frac
{\sum_{u \in \partial i} \sum_k \omega_{ui}(k,\ell)}
{\sum_{u \in \partial i} \sum_{k\ell} \omega_{ui}(k,\ell)} 
= \frac{\sum_{u \in \partial i} \sum_k \omega_{ui}(k,\ell)}{d_i} \, ,
\end{equation*}
where $d_i$ is the degree of item $i$.  This completes the derivation of~\eqref{eq:upd-theta} and~\eqref{eq:upd-eta}.  Finally, including a 
Lagrange multiplier for~\eqref{eq:norm-pr}, we have
\begin{equation*}
%\label{eq:update-pr}
p_{k\ell}(r) = \frac
{\sum_{(u,i) \in R^O | r_{ui}=r} \omega_{ui}(k,\ell)}
{\sum_{(u,i) \in R^O} \omega_{ui}(k,\ell)} \, ,
\end{equation*}
completing the derivation of~\eqref{eq:upd-pr}.

\section{Datasets}
\begin{table*}
\caption{Dataset characteristics. The total number of possible ratings is different for each dataset; ratings are in a scale from 1 to 5 in all datasets for the two dating agency datasets, which have a rating scale from 1 to 10.  Ratings are integers except for the Movielens 10M dataset which allows half-integer values.  Note that, in the latter case we expect a smaller MAE than if only integer values were allowed.  
%we had the 10 ratings in a scale from 1 to 10 with a precision of 1.  
All datasets have millions of ratings except for MovieLens 100K and Yahoo! Songs.  The average percentage of cold start cases is taken over all 5 test sets in the five-fold cross-validation experiment.}
\begin{tabular}{@{\vrule height 10.5pt depth0pt  width0pt}l|c|r|r|r|c}
\hline
Dataset &Ratings scale $S$&\#Users &\#Items &\#Ratings & Average cold start (\%) \\ 
%& & & & & cold start (\%)\\
\hline
MovieLens 100K &$\{1,2,3,4,5\}$ &943 & 1,682 & 100,000 &  0.17\%\\
%MovieLens& & & & & \\
MovieLens 10M & $\{0.5,1, 1.5, \ldots ,5\}$& 71,567 & 65,133 & 10,000,000 & 0.0015\% \\
%MovieLens& & & & & \\
Yahoo! Songs & $\{1,2,3,4,5\}$  & 15,400 & 1,000 & 311,700 & - \\
M-W dating agency&$\{1,2, \ldots,10\}$ & 220,970  & 135,359 & 4,852,455 & 0.31\% \\
%dating agency& & & & & \\
W-M dating agency&$\{1,2, \ldots,10\}$ & 135,359 &220,970 & 10,804,040 &  0.625\% \\
%dating agency& & & & & \\
Amazon book & $\{1,2,3,4,5\}$ & 73,091  & 539,145 & 4,505,893 & 6.7\% \\
\hline
\end{tabular}
\label{t-data}
\end{table*}
We perform experiments on six different datasets: the MovieLens 100K and 10M datasets (\text{movielens.umn.edu}), Yahoo! Songs (\text{research.yahoo.com/Academic\_Relations}, \text{ydata-ymusic-user-artistratings-v1\_0}), Amazon books (\text{jmcauley.ucsd.edu/data/amazon/}), and the LibimSeTi.cz dating agency (\text{occamslab.com/petricek/data/}). We split the LibimSeTi.cz dataset into two datasets: women rating men (W-M) and men rating women (M-W).  We neglected the links of women rating women and men rating men; unfortunately these links constituted only 1\% of the dataset.  
%There also a total 1\% of women rating women and men rating men that we neglected. 
In Table~\ref{t-data}, we show the characteristics of each dataset in terms of the scale of ratings $S$, the total number of users, the total number of items, the number of ratings and the average percentage of cold start cases.  
The MovieLens 100K dataset also provides demographic information for the users, namely the age in years and gender.

\section{Benchmark algorithms}

\textbf{Naive model} 
As a baseline for comparison, we consider a naive model.  Its prediction for a rating $r_{ui}$ is simply the average of $i$'s observed ratings,
\begin{equation}
r_{ui} = \frac{1}{d_i} \sum_{u' \in \partial_i} r_{u'i} \, . 
%\frac{\sum_{u' \in U_i} r_{u'i}}{|U_i|},
\end{equation}
%where $U_i$ are the users that rate item $i$, and $|U_i|$ are the number of these users.$\sum_k \theta_{uk}\theta_{vk}/(|\theta_u|_2 |\theta_v|_2)$

\textbf{Item-item} 
The item-item algorithm uses the cosine similarity between items, based on the $N$-dimensional vectors of ratings they have received, adjusted to remove user biases towards higher or lower ratings~\cite{sarwar01}.  The cosine similarity of items $i$ and $j$ is then $\cos (r_i,r_j) = \sum_u^N r_{iu} r_{ju} / (|r_i|_2 |r_j|_2)$.  
%vector . In this model, each item has a rating vector $\overrightarrow{i}$ which corresponds to the users that have rated this item, then the similarity between two items $sim(i,j)$ is the cosine between the item rating vectors as
%\begin{equation}
%\mathrm{sim}(i,j)=\frac{\overrightarrow{i} \cdot \overrightarrow{j}}{\|\overrightarrow{i}\|_2 * \|\overrightarrow{j}\|_2}.
%\label{eq:itemsim}
%\end{equation}
The predicted rating $r_{ui}$ is the similarity-weighted average of the $k$ closest neighbors of $i$ that user $u$ has rated. We use the default, optimized implementation of the algorithm in LensKit~\cite{ekstrand11} with $k=50$.

\textbf{Matrix factorization} 
One of the most widely used recommendation algorithms is matrix factorization (MF)~\cite{koren09,paterek07}.  Like the block model, the intuition behind matrix factorization is that there should be some latent features that determine how a user rates an item.  However, it uses linear algebra to reduce the dimensionality of the problem.  
%The matrix factorization method based on singular value decomposition (SVD) works as follows \cite{koren09}. 
%Instead of thinking that all ratings on the system are independent, it assumes that there are generalities that guide how users rate on items. In practice this is modelled in MF as free features, shared by users and items, such as the problem is dimensionally reduced. 
Specifically, it assumes that the matrix of ratings $R$ (with $N$ rows and $M$ columns) is of rank $k$, in which case it can be written $R=PQ$ where $P$ is a $N \times k$ matrix and $Q$ is a $k \times M$ matrix.  If we denote the rows of matrix $P$ as $p_u$ and the columns of $Q$ as $q_i$, then individual ratings are inner products $r_{ui}=p_{u} \cdot q_i$.

%; if we assume this noise is Gaussian, then the most-likely $P$ and $Q$ are given by the singular value decomposition.

%To solve this factorization problem we choose the singular value decomposition (SVD) method  \cite{paterek07}. This method find $P$ and $Q$ which product minimized the error with the original ratings matrix $R$ (specifically the means squared error). In the LensKit implementation of the algorithm we used for the predictions, this problem is numerically solved by stochastic gradient descent algorithm \cite{gardner03,koren09}. For the LensKit implementation we set $K=50$ and a learning rate of $0.002$ as suggested in Ref.~\cite{ekstrand11}.

%For the purpose of making recommendations, it is convenient to pose the decomposition problem as an optimization one; indeed, one can prove that $P$ and $Q$ are the solution of
%\begin{equation}
%\{p_u,q_i\}=\argmin_{\tilde{p_u}\tilde{q_i}}\sum_{ui}(r_{ui}-\tilde{p_u^t}\tilde{q_i})^2
%\end{equation}

We then assume that some noise and/or bias has been applied to $R$ to produce the observed ratings $R^O$.  For example, some users rate items higher than others, and some items are systematically highly rated. In order to take this into consideration, the unobserved ratings $r_{ui}$ are estimated using
\begin{equation}
r_{ui} = p_u \cdot q_i + \mu + b_u + b_i
\end{equation}
where $b_u$ and $b_i$ are the biases of users and items respectively and $\mu$ is the average rating in $R^O$. For the purpose of making recommendations, it is convenient to pose the decomposition problem as an optimization one; in particular, minimizing the $\ell_2$ error and applying a regularization term gives
\begin{equation}
\begin{split}
\{p_u,q_i\} 
&= 
\argmin_{\widetilde{p_u},\widetilde{q_i}}
\sum_{(u,i)\in R^O} \left[ (r_{ui}-\widetilde{p_u} \cdot \widetilde{q_i}-\mu-b_u-b_i)^2 \right.\\
& \left.+ \lambda (\lVert \widetilde{p_u}\rVert ^2 + \lVert \widetilde{q_i}\rVert ^2) \right] \, .
\end{split}
\end{equation}
where $\lambda$ is a regularization parameter.  As Funks originally proposed \cite{koren09} one can solve this problem numerically using stochastic gradient descent~\cite{gardner03}. We use the LensKit implementation of the algorithm, with $k=50$ and a learning rate of $0.002$ as suggested in Ref.~\cite{ekstrand11}.

\textbf{Stochastic block model} 
The stochastic block model (SBM)~\cite{holland83,nowicki01,guimera09} assumes that the probability that two nodes form a link between them, such as a relationship between actors in a social network, depends on what groups they belong to.   
%In this family of models, social actors are divided into groups and relationships between two actors are established depending solely on the groups to which they belong. 
Analogously, the SBM recommender algorithm~\cite{guimera12} assumes that the probability of a rating $r_{ui}$ of a user $u$ for an item $i$ depends on the groups $\sigma_u$, $\sigma_i$ to which they belong; unlike this paper, it assumes that each user or item belongs to a single group rather than a mixture.   
%The ratings in this case are treated as independent labels (ratings $1$ and ratings $2$ are not considered to be more related than ratings $1$ and $5$).  
%is mathematically sound because it 
It uses a Bayesian approach that deals rigorously with the uncertainty associated with the models that could potentially account for the observed ratings. Mathematically, the problem is to estimate $p(r_{ui}=r|R^O)$ that the unobserved rating of item $i$ by user $u$ is $r_{ui}=r$ given the observable ratings $R^O$.  This is an integral over all possible block models $M$,  
\begin{equation}
\label{eq:sbm-integral}
p(r_{ui}=r|R^O)=\int_{M} \: dM \:  p(r_{ui}=r|M) \: p(M|R^O) ,
\end{equation}
where $p(r_{ui}=r|M)$ is the probability that $r_{ui}=r$ if the ratings where actually generated using model $M$, and $p(M|R^O)$ is the probability of model $M$ given the observation (assuming for simplicity that all models $M$ are equally likely a priori).   
% Using Bayes theorem becomes
%\[
%p(r_{ui}|R^O)=\frac{\int_M \: dM \: p(r_ui=r|M)\: P(R^O|M)\: p(M)}{\int_{M'} \: dM' \: p(R^O|M') p(M')}
%\]
This integral is over the continuous and discrete parameters of the block model.  In particular, for each $r$ and each pair of groups $k, \ell$ we integrate over the continuous parameters $\Pr[r_{ui}=r | \sigma_u=k, \sigma_i=\ell] = p_{k\ell}(r)$; this part of the integral can be carried out analytically.  
%In practice, the models considered are the family of stochastic block models $M_{\mathrm{SBM}}$, where the of probability that a user rates an item will depends, exclusively, on the groups $\sigma_u$ and $\sigma_i$ to which the user and the item belong (one unique group for each user and also one group for each item), that is
%\begin{equation}
%p(r_{ui}=r)=p_{r(\sigma_u, \sigma_i).
%\end{equation}
However, the integral~\eqref{eq:sbm-integral} also averages over all assignments $\sigma$ of groups to users and items; this expectation is estimated by Metropolis-Hastings sampling.  Finally the prediction for each rating is the maximum-marginal estimate, 
\begin{equation}
r_{ui} = \argmax_r p_{\mathrm{SBM}}(r_{ui}=r|R^O),
\end{equation}
%Importantly, we obtain of the whole probability distribution for each rating (similarly to our MMSBM (Eq.~\ref{eq:model}).

% In contrast, recommender systems like MF and item-item give only the most probable rating. However, the SBM approach relies on Markov chain Monte Carlo sampling to make ratings' predictions and therefore does not scale to large datasets. Note that in the SBM the ratings are treated independently, without assuming linearity among them.

\begin{acknowledgments}
This work was supported by a James S. McDonnell Foundation Research Award (RG, MS), Spanish Ministerio de Economia y Comptetitividad (MINECO) Grants FIS2013-47532-C3 (AGL, RG, MSP) and FIS2015-71563-ERC (RG), European Union FET Grant 317532 (MULTIPLEX, RG, MSP), the John Templeton Foundation (CM) and the ARO under contract W911NF-12-R-0012 (CM).
\end{acknowledgments}

%\bibliography{./References/ref-database,add-refs}

%

\end{document}